\documentclass[aip,reprint]{revtex4-1} 
\usepackage{rotate}
\usepackage{rotating}
\usepackage{graphicx}
\usepackage{wrapfig}
\usepackage{comment}
\usepackage[caption=false]{subfig}
\usepackage{longtable}
\usepackage{appendix}
\usepackage{color}
\usepackage{setspace}
\usepackage{array}
\usepackage{siunitx}
\usepackage[sc]{mathpazo}
\usepackage[T1]{fontenc}

\usepackage{amsmath,amssymb}

\newcommand{\calder}{CALDER}
\newcommand{\cobalt}{$^{57}$Co}
\newcommand{\SubstrateThickness}{300\,$\mu$m}

\newcommand{\NoiseGlobalResolution}{$154\pm7$\,eV}
\newcommand{\NoiseLorenzo}{$82\pm4$\,eV}
\providecommand*{\un}[1]{\ensuremath{\mathrm{~#1}}}

\begin{document}

\title{High sensitivity phonon-mediated kinetic inductance detector with combined amplitude and phase read-out}

\author{L.~Cardani}
\email{laura.cardani@roma1.infn.it}
\affiliation{INFN - Sezione di Roma, Piazzale Aldo Moro 2, 00185, Roma - Italy}
\affiliation{Physics Department - Princeton University, Washington Road, 08544, Princeton - NJ, USA}
\author{N.~Casali}
\affiliation{INFN - Sezione di Roma, Piazzale Aldo Moro 2, 00185, Roma - Italy}
\author{I.~Colantoni}
\affiliation{Istituto di Fotonica e Nanotecnologie - CNR, Via Cineto Romano 42, 00156, Roma - Italy}
\author{A.~Cruciani}
\affiliation{INFN - Sezione di Roma, Piazzale Aldo Moro 2, 00185, Roma - Italy}
\author{F.~Bellini}
\affiliation{INFN - Sezione di Roma, Piazzale Aldo Moro 2, 00185, Roma - Italy}
\affiliation{Dipartimento di Fisica - Sapienza Universit\`{a} di Roma, Piazzale Aldo Moro 2, 00185, Roma - Italy}
\author{M.G.~Castellano}
\affiliation{Istituto di Fotonica e Nanotecnologie - CNR, Via Cineto Romano 42, 00156, Roma - Italy}
\author{C.~Cosmelli}
\affiliation{INFN - Sezione di Roma, Piazzale Aldo Moro 2, 00185, Roma - Italy}
\affiliation{Dipartimento di Fisica - Sapienza Universit\`{a} di Roma, Piazzale Aldo Moro 2, 00185, Roma - Italy}
\author{A.~D'Addabbo}
\affiliation{INFN - Laboratori Nazionali del Gran Sasso (LNGS), Via Giovanni Acitelli 22, 67010, Assergi (AQ) - Italy}
\author{S.~Di Domizio}
\affiliation{Dipartimento di Fisica - Universit\`{a} degli Studi di Genova, Via Dodecaneso 33, 16146, Genova - Italy}
\affiliation{INFN - Sezione di Genova, Via Dodecaneso 33, 16146, Genova - Italy}
\author{M.~Martinez}
\affiliation{INFN - Sezione di Roma, Piazzale Aldo Moro 2, 00185, Roma - Italy}
\affiliation{Dipartimento di Fisica - Sapienza Universit\`{a} di Roma, Piazzale Aldo Moro 2, 00185, Roma - Italy}
\author{C.~Tomei}
\affiliation{INFN - Sezione di Roma, Piazzale Aldo Moro 2, 00185, Roma - Italy}
\author{M.~Vignati}
\affiliation{INFN - Sezione di Roma, Piazzale Aldo Moro 2, 00185, Roma - Italy}

\begin{abstract}
Developing wide-area cryogenic light detectors with baseline resolution better than 20\,eV is one of the priorities of next generation bolometric experiments searching for rare interactions, 
as the simultaneous read-out of the light and heat signals enables background suppression through particle identification.
Among the proposed technological approaches for the phonon sensor, the naturally-multiplexed Kinetic Inductance Detectors (KIDs) stand out for their excellent intrinsic energy resolution and reproducibility. 
The potential of this technique was proved by the CALDER project, that reached a baseline resolution of \NoiseGlobalResolution\ RMS by sampling a 2$\times$2\,cm$^2$ Silicon substrate with 4 Aluminum KIDs.
In this paper we present a prototype of Aluminum KID with improved geometry and quality factor. The design improvement, as well as the combined analysis of amplitude and phase signals, allowed to reach a baseline resolution of \NoiseLorenzo\  by sampling the same substrate with a single Aluminum KID.
\end{abstract}

\maketitle

Bolometric experiments searching for rare events, such as neutrino-less double beta decay or dark matter interactions, are now focusing on the development of cryogenic light detectors to enable background suppression exploiting the different light yield of different particles\cite{Wang:2015taa}.
The ideal light detector should provide excellent energy resolution ($<$20\,eV), wide active surface (5$\times$5\,cm$^2$), reliable and reproducible behavior, and the possibility of operating hundreds/thousands of channels.
None of the existing technologies~\cite{Casali:2015gya,Casali:2014vvt,Beeman:2011yc,Gironi:2016nae,Willers:2014eoa,Angloher2016} is ready to fulfill all these requirements without further R$\&$D. 
Since most of the proposed detectors are limited by the number of channels that can be easily installed and operated, the \calder\ project\cite{CalderWhitePaper} aims to develop a light detector starting from devices that are naturally multiplexed, such as the Kinetic Inductance Detectors (KIDs)\cite{Day:2003fk}.
Thanks to the high sensitivity and to the multiplexed read-out, KIDs have been proposed in several physics sectors, like photon detection, astronomy\cite{Day:2003fk,Monfardini2011,Mazin:2013wvi}, search for dark matter interactions\cite{Golwala2008,moore2}, and for the read-out of transition-edge sensors arrays\cite{Noroozian:2013,Giachero:2016ehv}.
KIDs show all the desirable features for an innovative light detector with the exception of a wide active surface:
macro-bolometers used by experiments such as CUORE\cite{Artusa:2014lgv} and CUPID-0\cite{Beeman:2013sba,Artusa:2016maw} are characterized by surfaces of several cm$^2$, while typical KIDs sizes barely reach few mm$^2$.
This limit can be overcome by following the phonon mediated approach\cite{swenson,moore2}: photons are coupled to the KIDs through a large insulating substrate, that converts them into phonons.
The athermal phonons that are not thermalized or lost through the substrate supports, reach the superconductor and break Cooper pairs, giving rise to the signal.

The first CALDER prototype\cite{Cardani:2015tqa}, obtained by depositing four 40\,nm thick Al KIDs on a 2$\times$2\,cm$^2$, \SubstrateThickness\ thick Si substrate, reached a combined baseline resolution of \NoiseGlobalResolution\ and a single KID absorption efficiency of 3.1 -- 6.1$\%$ depending on the position of the source.
In this paper we present a resonator design that allows to improve the KID efficiency up to 7.4 -- 9.4$\%$, and to reach a baseline resolution of \NoiseLorenzo\ with a single KID on a similar substrate.

\begin{figure}[htbp]
 \includegraphics[width=.30\textwidth, natwidth=1005, natheight=787]{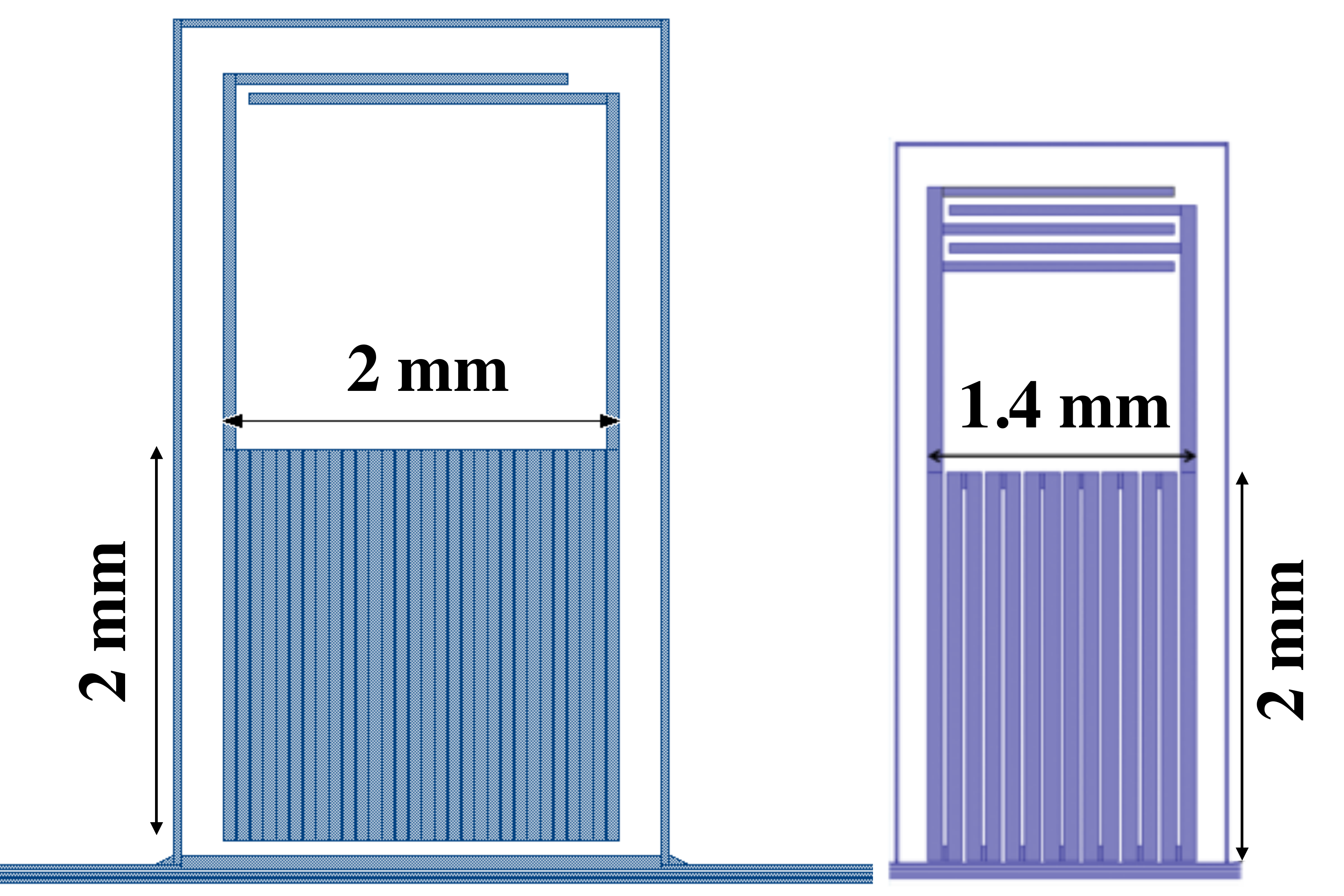}
  \caption{\label{fig:sketch} Left: diagram of the pixel described in this paper. The inductor (30 strips of $62.5\,\mu\rm m \times 2\un{mm}$) features an active area of 3.75\,mm$^2$ (4.0\,mm$^2$ including the active region that connects the inductor to the capacitor). To contain the geometrical inductance, we used a gap of $5\,\mu\rm m$ between the meanders and we closed the circuit using a capacitor made by  only 2 fingers. Right: diagram of the pixel described in  Ref.\cite{Cardani:2015tqa}, reproduced from Appl. Phys. Lett. 107, 093508 (2015), with the permission of AIP Publishing. The active area of the inductive meander (2.4\un{mm^2}) consists of 14 connected strips of $80\,\mu\rm m\times 2\un{mm}$, with a gap of $20\,\mu\rm m$.
}
\end{figure}

To improve the detector resolution we tested KIDs with different geometries on 2$\times$2\,cm$^2$, 380\,$\mu$m thick, high resistivity ($>$10\,k$\Omega\times$cm) Si(100) substrates. We deposited a single KID on each substrate in order to characterize the detector response in absence of cross-talk or competition among pixels in the absorption of the propagating phonons. As discussed later in the text, all the tested prototypes featured an excess low-frequency noise constistent with what observed in our first prototype\cite{Cardani:2015tqa}. Therefore to improve the signal-to-ratio (SNR) we tried to increase the signal. First, we raised the quality factor of the resonator (1/Q = 1/Q$_c$ + 1/Q$_i$): Q$_c$ was raised up from 6-35$\times 10^3$ to $\sim150\times 10^3$ by design and,  to ensure a high Q$_i$, we used a 60\,nm thick film, since thicker films are generally characterized by a better quality of the superconductor. Then, we enlarged the active area of the KID from 2.4 to 4.0\,mm$^2$, in order to increase the fraction of phonons that can be collected before being lost in the substrate. A comparison of the improved design with the one described in Ref.\cite{Cardani:2015tqa} is shown in Fig.~\ref{fig:sketch}.

The resonator was patterned by electron beam lithography on a single Al film deposited using electron-gun evaporator (more details on the design and fabrication processes can be found in Refs.\cite{Colantoni2016,Colantoni:2016tpk}).
The chip was mounted in a copper holder using PTFE supports with total contact area of about 3\,mm$^2$, and connected to SMA read-out by ultrasonic wire bonding.
The detector was anchored to the coldest point of a $^3$He/$^4$He dilution refrigerator with base temperature of 10\,mK.
The output signal was fed into a CITLF4 SiGe cryogenic low noise amplifier\cite{amply} with T$_N\sim$7\,K. Details about the room-temperature electronics and acquisition can be found in Refs.\cite{CalderWhitePaper,Bourrion:2011gi,Bourrion:2013ifa}.

\begin{figure}[htbp]
 \includegraphics[width=.47\textwidth, natwidth=362, natheight=567]{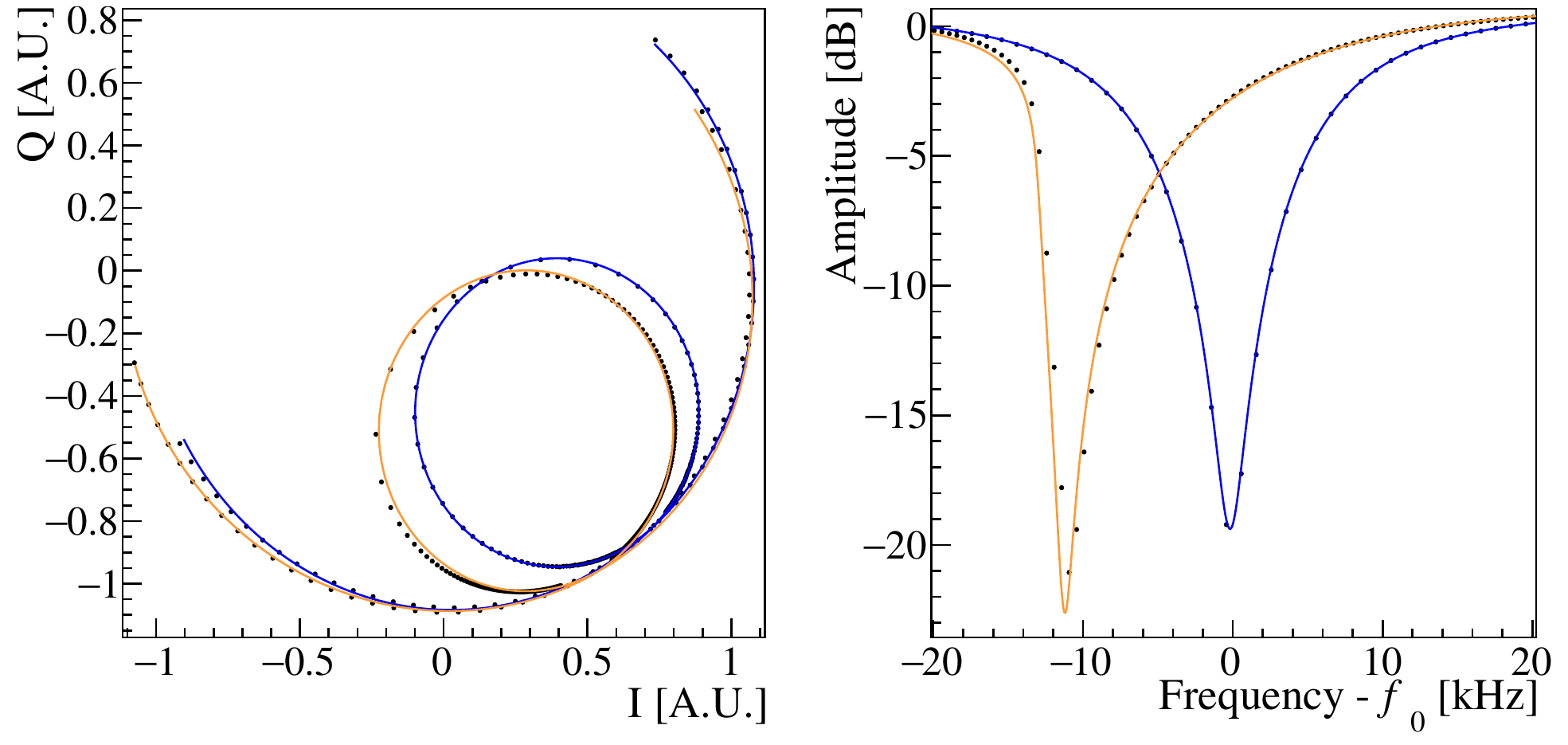}
  \caption{\label{fig:fit resonances} Data (dots) and fit (line) of the complex transmission $S_{21}$ past the resonator for a frequency sweep at low microwave power (-97\,dBm, blue line) and in the working point (-62\,dBm, orange). Left: imaginary part (Q) as a function of the real part (I) of $S_{21}$. Right: amplitude of $S_{21}$ as a function of the frequency. Data were scaled to fit in the same plot.}
\end{figure}

The resonance parameters (Q, Q$_c$, Q$_i$,  $f_0$) were derived by fitting the complex transmission $S_{21}$ measured in a frequency sweep around the resonance using the model described in Ref\cite{Casali:2015bhk} (Fig.~\ref{fig:fit resonances}).
The large Q$_c=156\times10^3$ limits the accuracy on the evaluation of Q$_i$, which is however very high: at low microwave power ($P_{\mu w}$), where Q$_i$ saturates, we obtain Q$_i > 2\times10^6$ and Q$\ =147\times10^3$.
To test the reproducibility of this device, we fabricated and measured another prototype with the same design, obtaining similar values of Q$_c$ and Q$_i$.

We derived the fraction $\alpha$ of kinetic inductance to the total inductance. 
We measured the shift of the resonant frequency and of the internal quality factor as a function of the temperature between 10 and 400\,mK.
We fitted the obtained data to the Mattis Bardeen theory using the approximated formulas derived by Gao et al.\cite{GAOvsMattisBardeen}, in which the only free parameters are $\alpha$ and the superconductor gap 2$\Delta_0$. Since these parameters are found to be highly correlated in the fit, 
we performed a direct and independent measurement of the transition temperature to infer $\Delta_0$. We obtained $T_c=1.18\pm0.02$\,K, corresponding to  $\Delta_0=179\pm3\,\mu$eV. 
Fixing $\Delta_0$ in the fit, we derived $\alpha=2.54\pm0.09^{stat}\pm0.26^{syst}\,\%$ from the fit of the shift of the resonant frequency.
This value is in good agreement with the one by obtained by fitting the shift of the inverse internal quality factor with temperature: $\alpha_Q=3.07\pm0.19^{stat}\pm0.30^{syst}$.

\begin{figure}[htbp]
 \includegraphics[width=.47\textwidth, natwidth=567, natheight=363]{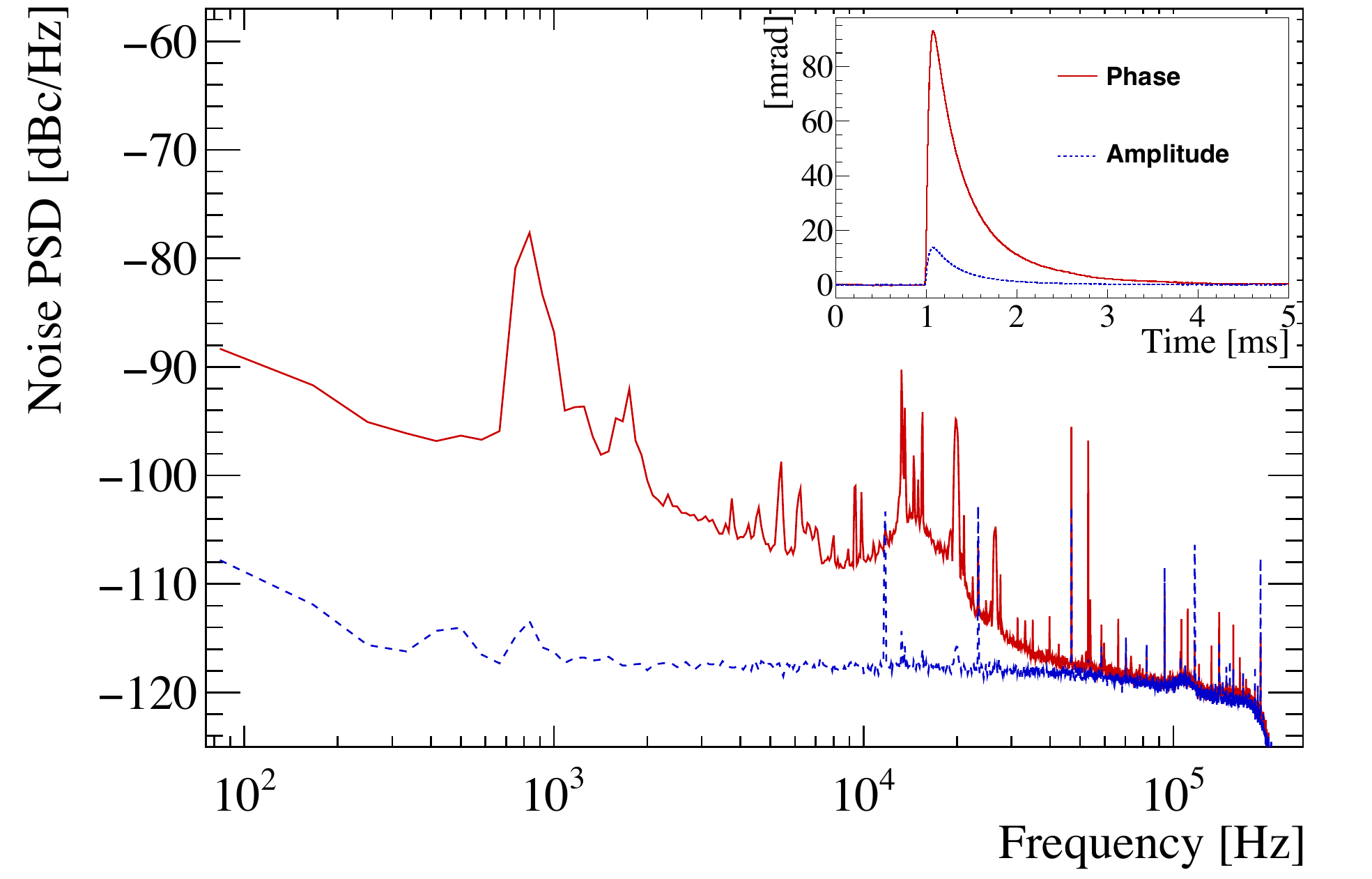}
  \caption{\label{fig:PulsesAndNoise}Noise power spectral density of $\delta A$ (dotted blue line) and $\delta\phi$ (continuous red line).  Inset: $\delta A$ and $\delta\phi$ response to 15.5\,keV optical pulses. Data were acquired by with $P_{\mu w}$ optimizing the SNR (-62\,dBm). The peak around 800 Hz in the phase noise comes from the cryogenic environment but it does not affect the energy resolution significantly.}
\end{figure}

We acquired 12\,ms long time windows with sampling frequency of 500\,kHz for the real ($I$) and imaginary ($Q$) parts of $S_{21}$.
$I$ and $Q$ were then converted during the off-line analysis into amplitude ($\delta A$) and phase ($\delta\phi$) variations relative to the center of the resonance circle.
The typical response to pulses with nominal energy of 15.5\,keV, obtained by averaging hundreds of events to suppress the random noise contribution, is reported in the inset of Fig.~\ref{fig:PulsesAndNoise}.
Pulses were produced by fast burst of photons emitted by a room-temperature 400\,nm LED, coupled to an optical fiber facing the backside of the chip to prevent direct illumination.
The optical system was calibrated at room temperature using a photomultiplier and correcting the results with a Monte Carlo simulation that accounted for the geometry of the final set-up (including the reflectivity of the materials). The room-temperature calibration was cross-checked at lower temperatures using a \cobalt\  X-rays source (main peaks at 6.4 and 14.4\,keV).\\

Thanks to the high resonator Q, we obtained a signal height of $\sim$5.8\,mrad/keV in $\delta\phi$ and 0.6\,mrad/keV in $\delta A$, about a factor 6 larger with respect to $\delta\phi$ and $\delta A$ obtained with our previous prototype\cite{Cardani:2015tqa}. We observe an improvement of the SNR in the phase direction that however does not scale linearly with $Q$, since we also observe an increase of the low frequency noise.
On the other hand, the amplitude noise is consistent with the amplifier temperature and much lower than the phase one (Fig.~\ref{fig:PulsesAndNoise}). For this reason, even if the amplitude signals are smaller than the phase ones, the SNR ratios are similar.  
We tried to reduce the low-frequency phase noise  by changing the room-temperature readout system, by driving the electronics with a Rubidium-referenced clock,
and by testing different groundings on the whole electronics and cryostat setup. None of these attempts succeeded, and therefore we ascribe the noise origin to the chip. 
Part of this noise could be ascribed to TLS, as suggested by the behavior of the resonance at different microwave powers (see Fig.~\ref{fig:fit resonances}). Nevertheless, the shape of the measured noise power spectrum can not be entirely described by this noise source. 
We plan to investigate it in the future by further changing the spacing and the width of the inductive meander of the resonator.

Pulses and noise windows of $\delta A$ and $\delta\phi$ were processed with the matched filter\cite{Gatti:1986cw}. To further improve the SNR
 we combined them with a 2D matched filter:
\begin{equation}
\vec {H}^{T}(\omega) = h\, \vec{S}^\dag(\omega)  N^{-1}(\omega),
\end{equation}
where $h$ is a normalization constant, $S$ is the vector of the $\delta A$ and $\delta\phi$  template signals, and $N$ is the covariance matrix of the noise.
We note that, compared to the combination proposed by Gao~\cite{gaoPhD}, this filter includes the noise correlation which may be significant in case of generation-recombination or readout noise. In our case, however, the correlation in the signal band is almost negligible and the gain of the combination arises from the comparable SNR of $\delta A$ and $\delta\phi$.

To determine the best microwave power, we studied the SNR after the matched filter, 
and chose $P_{\mu w}$ that allowed to maximize this parameter:  $P_{\mu w}^{opt}=-62$\,dBm.
Finally, we calibrated the energy scale and checked the linearity of the detector response by means of optical pulses with energy ranging from 3 to 31\,keV. \\

\begin{figure}[htbp]
 \includegraphics[width=.47\textwidth, natwidth=362, natheight=567]{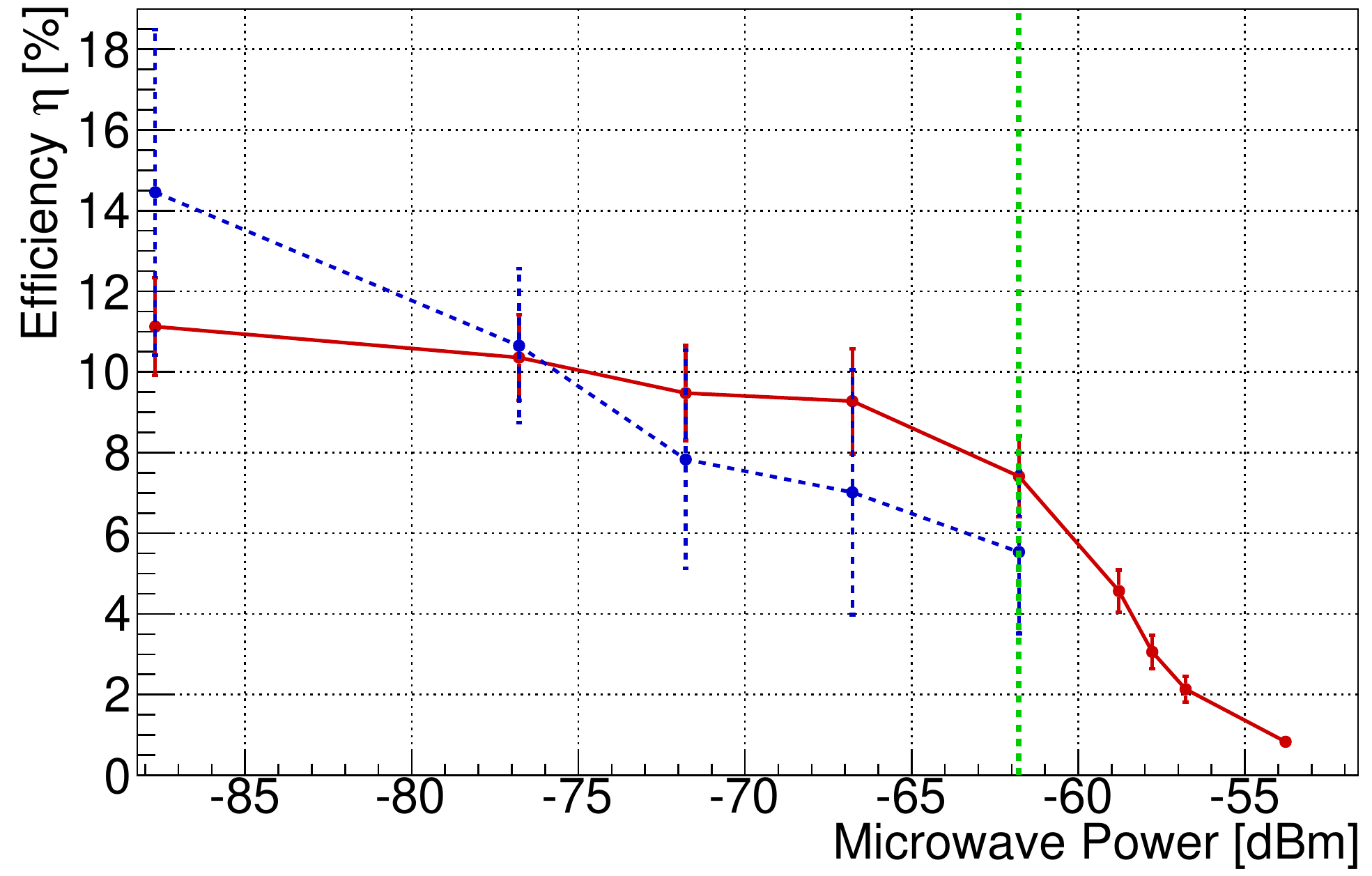}
 \caption{\label{fig:gain}Efficiency of the detector evaluated using $\delta\phi$ (red, continuous line) and $\delta A$ (blue, dotted line). The optical source was placed far from the KID. The vertical bar indicates $P_{\mu w}^{opt}$.}
\end{figure}

The enlargement of the KID geometry allows to improve the detector efficiency and, therefore, the energy resolution. The efficiency $\eta$ can be computed by comparing the nominal energy with the energy absorbed by the resonator:
$E  = \frac{1}{\eta} E_{absorbed} = \frac{1}{\eta} \Delta_0 \delta n_{qp}$
where $\delta n_{qp}$ is the variation of the number of quasiparticles. To compute $\delta n_{qp}$, we deepen
the analysis proposed by Moore et al.\cite{moore2} by considering the dependence of the KID response on its effective temperature 
and by extending the analysis also to the amplitude signal (in addition to the phase one).
In a simplified model that assumes a thermal quasiparticles distribution, phase and amplitude variations can be related to the energy through the following formulae:
\begin{align}
\label{eq:finalE_ampl}
E^{\delta A}     &= \frac{1}{\eta_{A}} \frac{N_0V\Delta_0^2}{\alpha S_1(f_0,T) Q} \delta A \quad 
E^{\delta\phi}  &=\frac{1}{\eta_{\phi}} \frac{N_0V\Delta_0^2}{\alpha S_2(f_0,T) Q} \delta\phi 
\end{align}
where $\eta_{A}$ and $\eta_{\phi}$ are the efficiencies calculated starting from $\delta A$ and $\delta\phi$ respectively,
N$_0$V is the single spin density of states at the Fermi level (N$_0=1.72\times10^{10}$\,eV$^{-1}\mu$m$^{-3}$ for Al) multiplied for the active volume of the resonator $V$, and $S_1{(f_0,T)}$/$S_2{(f_0,T)}$ are functions of $\Delta_0$, of the effective detector temperature (that depends on $P_{\mu w}$) and of the resonant frequency $f_0$\cite{moorePhD}.
For each power, we derived the effective temperature from the frequency shift and we computed the corresponding values of $S_1(f_0,T)$ and $S_2(f_0,T)$.
Substituting the obtained values in Eq.\eqref{eq:finalE_ampl} we computed $\eta_{\phi}$ and $\eta_{A}$ as a function of $P_{\mu w}$ (Fig.~\ref{fig:gain}).
In the power range of interest, $\eta_{A}$ and $\eta_{\phi}$ show differences lower than 35$\%$ even if we started from independent formulae for the calibration of phase and amplitude.

\begin{figure}[htbp]
 \includegraphics[width=.47\textwidth, natwidth=567, natheight=385]{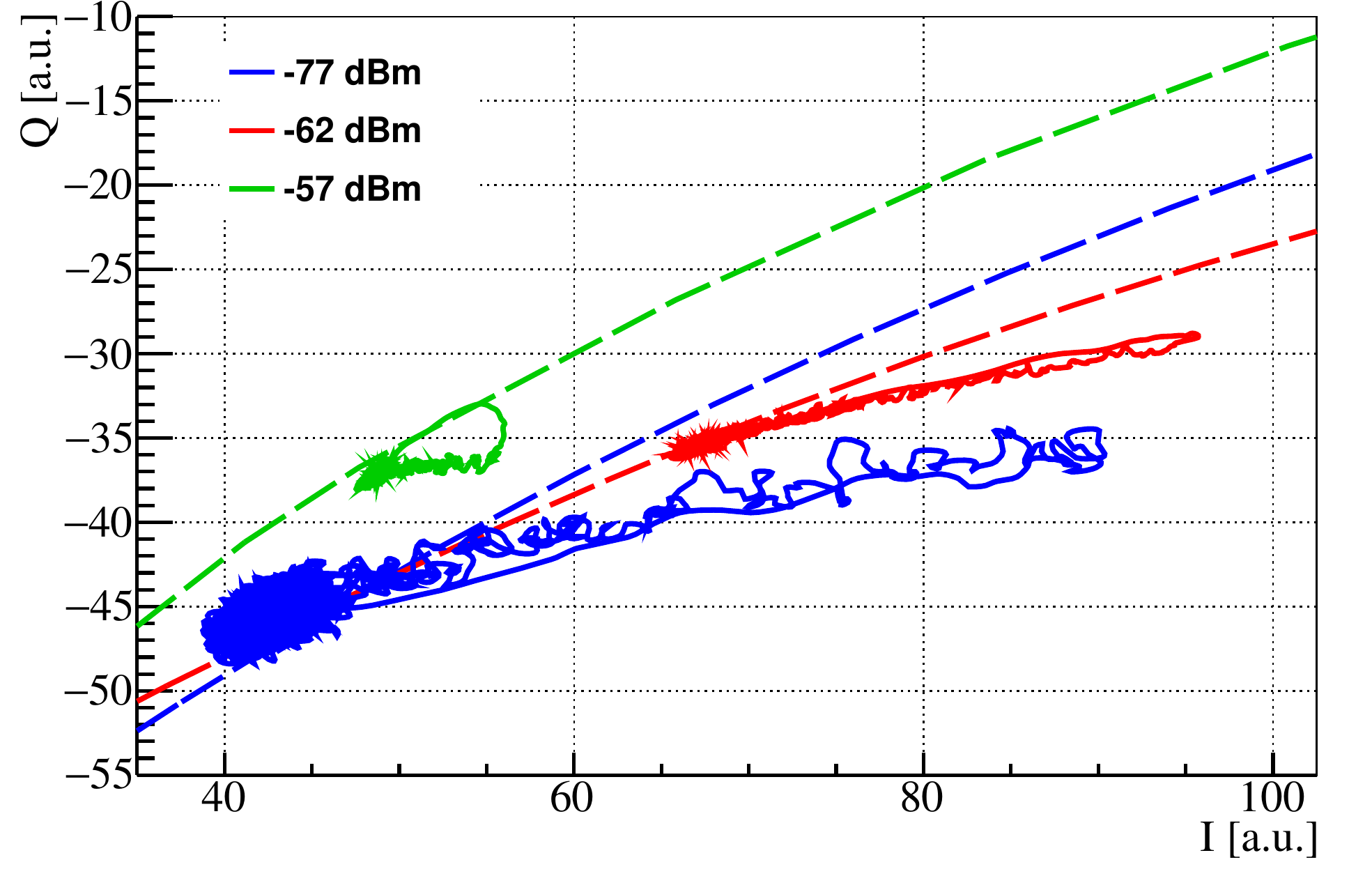}\\
 \caption{\label{fig:bananas}Response to 15.5\,keV pulses (line) along the resonance frequency scan (dotted line) for different microwave powers ($P_{\mu w}$). All curves were scaled by $\sqrt{P_{\mu w}}$. At high $P_{\mu w}$ the electrothermal feedback visibly affects the pulse trajectory.}
\end{figure}

To estimate the errors on $S_1(f_0,T)$ and $S_2(f_0,T)$ we studied how reasonable variations of the effective temperature affected the values assumed by the two functions.
$S_1(f_0,T)$ depends rather sharply on the effective detector temperature in the region of interest (T$<$250\,mK). 
As a consequence, even small temperature variations can lead to errors of about 30 -- 40$\%$, that dominate the uncertainty on $\eta_{A}$. 
On the contrary, $S_2(f_0,T)$ is a slow function of the temperature, and thus it introduces a smaller uncertainty on $\eta_{\phi}$ ($<$9$\%$). 
For $P_{\mu w}$ larger than -62\,dBm the resonance becomes too asymmetric to extract the frequency shift and, thus, the effective temperature. Errors on $S_1(f_0,T)$ become too large to study $\eta_{A}$. On the contrary, assuming a constant value for $S_2(f_0,T)$ allows to keep the uncertainty on $\eta_{\phi}$ lower than 10$\%$. Using this estimator, we observe that the efficiency decreases with the microwave power. 
For $P_{\mu w}$ higher than -62\,dBm the resonance shows a bifurcation\cite{swenson2013}, which affects also the pulses development (Fig~\ref{fig:bananas}). Thus, the relation between $\delta A$/$\delta\phi$ and $\delta n_{qp}$ may no longer be described by this simple model and, as a consequence, also our evaluation of $\eta$ may not be accurate at higher powers.

The detector efficiency was inferred from two measurements: by illuminating a $\sim$6\,mm diameter spot as far as possible from the resonator, and by placing the source below the KID (always on the opposite face of the substrate). The first configuration was chosen to be conservative, as placing the source far from the KID decreases the phonon collection efficiency.
At $P_{\mu w}^{opt}$, we obtained $\eta_{\phi} = (7.4\pm0.9)\%$ with the optical source far from the KID, and $\eta_{\phi}=(9.4\pm1.1)\%$ with the source below the KID. 
Thus, the geometry presented in this paper allows to improve the efficiency, that for the detectors reported in Ref.\cite{Cardani:2015tqa} reached the maximum value of 6.1$\%$, when placing the source as close as possible to the resonator. Since the energy resolution scales linearly with the detector efficiency, we expect a similar improvement also in the sensitivity.
We highlight that working at powers lower than $P_{\mu w}^{opt}$ would allow to further increase the efficiency, as the signal height becomes larger.
Nevertheless, at lower powers also the noise of the detector increases, and overall the energy resolution is worst.

In principle, the energy resolution depends also on the quasiparticles recombination time $\tau_{qp}$ which is expected to decrease with  microwave power~\cite{Cruciani:2016moq,visser2014}.
In our case, for $P_{\mu w}$ lower than -77\,dBm, $\tau_{qp}$ is almost constant around 220\,$\mu$s because the power absorbed from the optical  pulses  is  higher than the one absorbed from the microwave.
At  higher power, likely for the presence of an electrothermal feedback~\cite{Thomas2015}, the pulse trailing and leading edges  follow  different trajectories in the IQ plane (Fig.~\ref{fig:bananas}), and the pulse decay time can be no longer interpreted as $\tau_{qp}$. In this power region, however, the efficiency drop is more significant and drives the loss in resolution.

\begin{figure}[htbp]
 \includegraphics[width=.47\textwidth, natwidth=362, natheight=567]{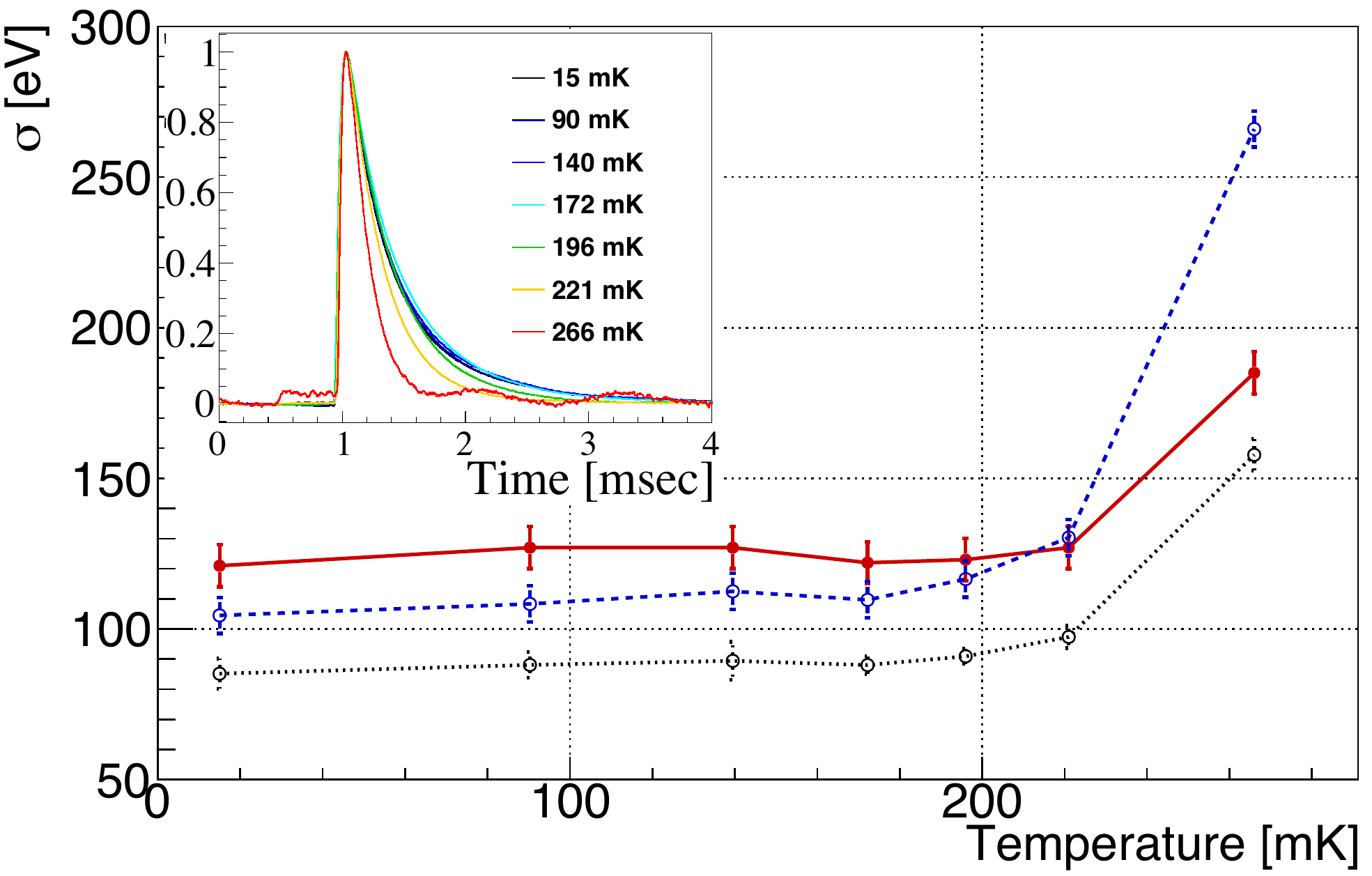}\\
 \caption{\label{fig:ResolutionVsT}Baseline RMS of $\delta\phi$ (continuous, red), $\delta A$ (dashed, blue) and their combination (dotted, black) as a function of temperature. Inset: $\delta\phi$ pulses at different temperatures; the amplitude is scaled to highlight the time-development.}
\end{figure}

We evaluate the sensitivity at $P_{\mu w}^{opt}$ from the baseline RMS $\sigma_{baseline}$
first for $\delta A$ and $\delta\phi$ to compare with other prototypes, and then for their combination with the 2D matched filter.
The results are $115\pm7$\,eV and $105\pm6$\,eV for $\delta A$ and $\delta\phi$, respectively, and  $82\pm4$\,eV for their combination. 
This latter value is a factor 2 better with respect to Ref.~\cite{Cardani:2015tqa} and with a single resonator instead of four.
This value is rather conservative, as the source was placed as far as possible from the KID.
We made a measurement with a source placed below the KID, obtaining a combined baseline resolution of $73\pm4$\,eV.
Finally, we proved that the detector baseline resolution is not affected by the temperature up to 200\,mK (Fig.~\ref{fig:ResolutionVsT}).

This work was supported by  the European Research Council (FP7/2007-2013) under contract  CALDER no. 335359 and  by the Italian Ministry of Research under the  FIRB  contract no. RBFR1269SL. 

This work was supported by  the European Research Council (FP7/2007-2013) under contract  CALDER no. 335359
and  by the Italian Ministry of Research under the  FIRB  contract no. RBFR1269SL.

\bibliography{../../../calder}

\end{document}